\documentclass[]{spie}  
\usepackage[]{graphicx}
\usepackage{amsmath,amssymb,mathtools,url}
\title{\Large \bf Fast Linearized Coronagraph Optimizer (FALCO) \\ \large IV. Coronagraph design survey for obstructed and segmented apertures } 
\usepackage{array}
\usepackage{hyperref}
\newcolumntype{P}[1]{>{\centering\arraybackslash}p{#1}}

\author{
G.~Ruane\supit{a,$\dagger$}, A.~Riggs\supit{b}, C.~T.~Coker\supit{b,c}, S.~B.~Shaklan\supit{b}, E.~Sidick\supit{b}, D.~Mawet\supit{a,b}, J.~Jewell\supit{b}, K.~Balasubramanian\supit{b}, C.~C.~Stark\supit{d}, \\
\supit{a}Department of Astronomy, California Institute of Technology, 1200 E. California Blvd., Pasadena, CA 91125, USA \\
\supit{b}Jet Propulsion Laboratory, California Institute of Technology, 4800 Oak Grove Dr., Pasadena, CA 91109, USA\\
\supit{c}NASA Postdoctoral Program Fellow\\
\supit{d}Space Telescope Science Institute, 3700 San Martin Dr, Baltimore, MD 21218, USA\\
}

\authorinfo{$^\dagger$Send correspondence to gruane@caltech.edu}

\graphicspath{{Figures/}}
 
\begin{document} 
  \maketitle 

\begin{abstract}
Coronagraph instruments on future space telescopes will enable the direct detection and characterization of Earth-like exoplanets around Sun-like stars for the first time. The quest for the optimal optical coronagraph designs has made rapid progress in recent years thanks to the Segmented Coronagraph Design and Analysis (SCDA) initiative led by the Exoplanet Exploration Program at NASA's Jet Propulsion Laboratory. As a result, several types of high-performance designs have emerged that make use of dual deformable mirrors to (1) correct for optical aberrations and (2) suppress diffracted starlight from obstructions and discontinuities in the telescope pupil. However, the algorithms used to compute the optimal deformable mirror surface tend to be computationally intensive, prohibiting large scale design surveys. Here, we utilize the Fast Linearized Coronagraph Optimizer (FALCO), a tool that allows for rapid optimization of deformable mirror shapes, to explore trade-offs in coronagraph designs for obstructed and segmented space telescopes. We compare designs for representative shaped pupil Lyot and vortex coronagraphs, two of the most promising concepts for the LUVOIR space mission concept. We analyze the optical performance of each design, including their throughput and ability to passively suppress light from partially resolved stars in the presence of low-order aberrations. Our main result is that deformable mirror based apodization can sufficiently suppress diffraction from support struts and inter-segment gaps whose widths are on the order of $\sim$0.1\% of the primary mirror diameter to detect Earth-sized planets within a few tens of milliarcseconds from the star. 
\end{abstract}



\section{INTRODUCTION}
\label{sec:intro} 

One of the main goals of the Large UV Optical Infrared Surveyor (LUVOIR) mission concept is to find and characterize a diverse sample of exoplanets through direct imaging and spectroscopy\cite{Bolcar2018}. The coronagraph instrument requirements are selected to maximize the number of Earth-sized planets studied in the habitable zone of nearby solar type stars\cite{Stark2014,Stark2015}. Not only will LUVOIR potentially detect tens of such objects, it will also yield hundreds of super-Earths, sub-Neptunes, and gas giants as well as images of circumstellar disks in scattered light. This will be made possible by a coronagraph instrument sensitive to planet-to-star flux ratios on the order of $10^{-10}$ as close as tens of milliarcseconds from the star from the visible to the near-infrared\cite{Pueyo2017}.

Achieving a raw contrast of $10^{-10}$ on a telescope with a segmented primary mirror requires coronagraph masks that are specially tailored to suppress, or ``apodize," the complicated diffraction patterns from the gaps between mirror segments and potential obscurations due to an on-axis secondary mirror and its corresponding support structures. Apodization of the stellar point spread function (PSF) is either achieved by changing the pupil shape\cite{Kasdin2003}, introducing a pupil mask that imparts a grayscale pattern\cite{Soummer2003_APLC}, or shaping the beam with phase induced amplitude apodization using fixed beam shaping optics\cite{Guyon2005} or active deformable mirrors (DMs)\cite{Pueyo2013,Trauger2016}, or a combination of any of the above. Several methods for the active compensation of aperture discontinuities (ACAD) with DMs have been introduced, including computing geometric solutions\cite{Pueyo2013} and closed-loop wavefront control methods\cite{Mazoyer2018a,Mazoyer2018b} that make use of electric field conjugation\cite{Giveon2007}, stroke minimization\cite{Pueyo2009}, and expectation-maximization\cite{Jewell2017} algorithms.

The Fast Linearized Coronagraph Optimizer (FALCO)\cite{Riggs2018} is a new open-source software package that calculates the DM surface shapes that minimimize the amount of starlight in a predetermined region of the image creating a so-called ``dark hole." FALCO significantly reduces the amount of time needed for such calculations with respect to previous methods used by our team\cite{Sidick2018}. Here, we use FALCO to perform a coronagraph design study for potential LUVOIR telescope apertures with shaped pupil Lyot\cite{Zimmerman2016} and vortex coronagraphs\cite{Mawet2005,Foo2005,Ruane2018_JATIS}. 

In the following sections, we present five coronagraph designs for potential LUVOIR apertures and compare their performance in terms of raw contrast, throughput, and robustness to stellar size and low order aberrations. We also study the effect secondary mirror strut size and the width of the gaps between mirror segments.

\begin{figure}[t!]
    \centering
    \includegraphics[width=0.6\linewidth]{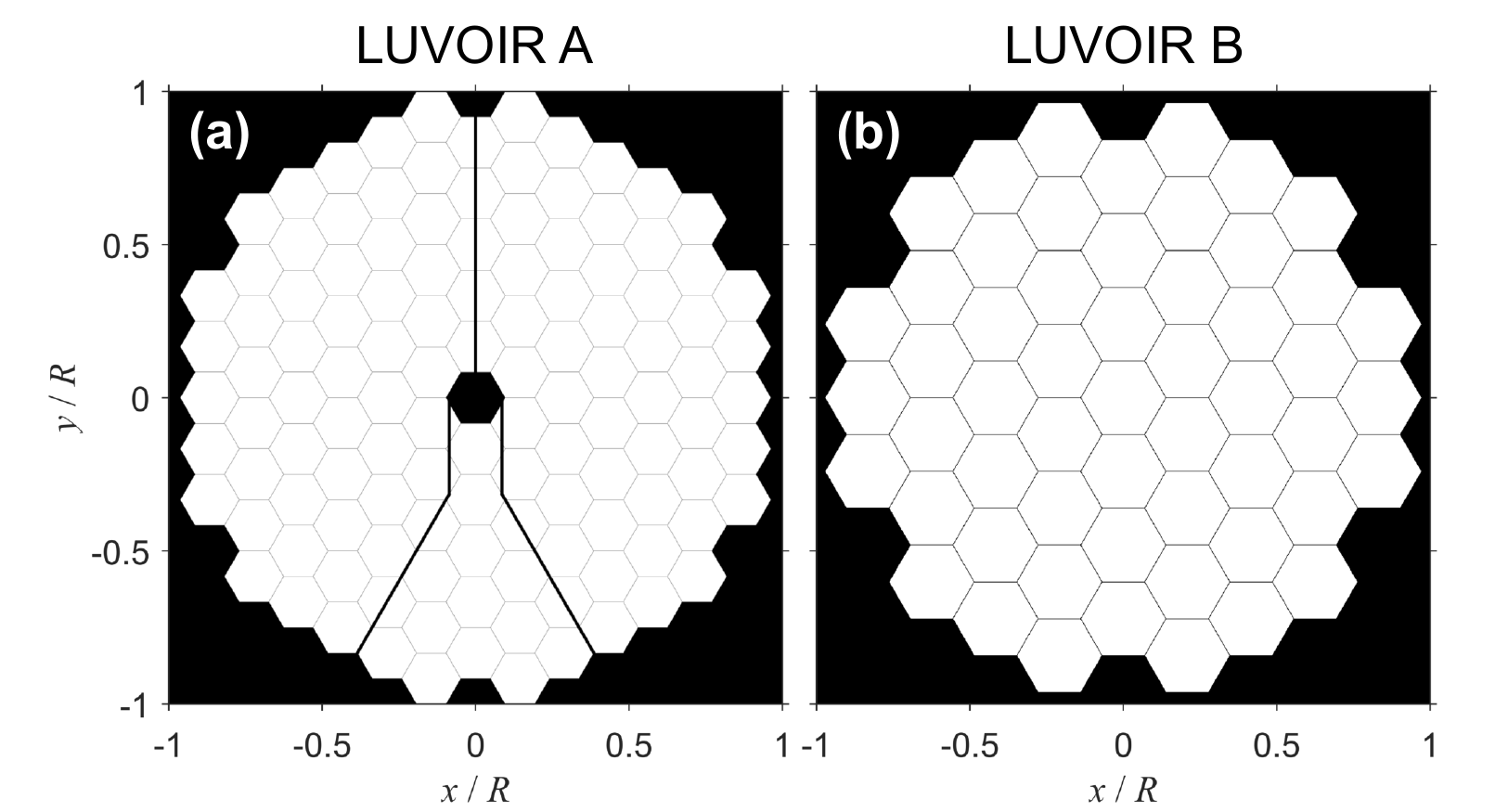}
    \caption{Potential apertures for LUVOIR. (a) LUVOIR~A is on-axis, centrally obscured, and segmented. (b) LUVOIR~B is off-axis and segmented.}
    \label{fig:aperture}
\end{figure}

\section{Telescopes and instrument}

Although the LUVOIR telescope design is still under development in preparation for the 2020 decadal survey, we analyze the performance of coronagraphs for the two potential architectures shown in Fig. \ref{fig:aperture}: an on-axis, centrally obscured aperture (LUVOIR~A) and an off-axis, unobscured aperture (LUVOIR~B). The primary mirror of LUVOIR~A is made up of 120 hexagonal segments with the central segment missing. The central obscuration has an circumscribed diameter that is 10\% of the inscribed diameter of the outer edge. LUVOIR~B, on the other hand, is made up of 55 segments and is otherwise unobscured. 

\begin{figure}[t]
    \centering
    \includegraphics[width=\linewidth]{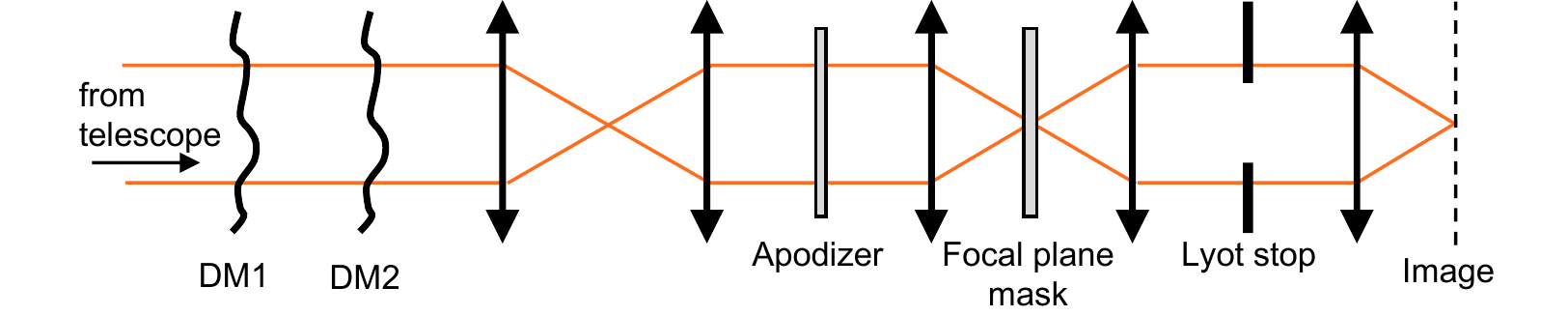}
    \caption{Schematic layout of a coronagraph with two deformable mirrors (DM1 and DM2), an apodizer mask, focal plane mask and Lyot stop. DM1, the apodizer, and the Lyot stop are in re-imaged pupil planes.}
    \label{fig:layout}
\end{figure}

Figure \ref{fig:layout} shows a schematic of the coronagraph instrument. We assume the coronagraph has two 64$\times$64-actuator deformable mirrors with 62 actuators across the circumscribed pupil diameter. DM1 is in the pupil plane while DM2 is at a distance of $R^2/(550\lambda)$ downstream, where $R$ is the pupil radius and $\lambda$ is the wavelength. Each actuator is represented by the default influence function distributed with the PROPER\footnote{\href{http://proper-library.sourceforge.net/}{http://proper-library.sourceforge.net/}} software package\cite{Krist2007,KristTDEM}. 


\section{Designs for LUVOIR A}

In this section, we present shaped pupil Lyot and vortex coronagraph designs for the LUVOIR~A aperture (Fig.~\ref{fig:aperture}a). In each case, we utilize an apodizer optimized to suppress the diffraction from the central obscuration and then compute the DM shapes needed to achieve a raw contrast of $<10^{-10}$ for the LUVOIR~A aperture, including the struts and segment gaps. 

\subsection{Shaped pupil Lyot coronagraph}

A shaped pupil Lyot coronagraph (SPLC) consists of a binary apodizer, an annular occulting mask in the intermediate focal plane, and annular Lyot stop that is undersized with respect to the re-imaged telescope pupil\cite{Zimmerman2016}. They are a type of apodized pupil Lyot coronagraph\cite{Soummer2003_APLC,Soummer2005_APLC,Soummer2009_APLCII,Soummer2011_APLCIII,NDiaye2015_APLCIV,NDiaye2016_APLCV}.

\subsubsection{1-D radial designs}

In Coker et al. (2018, these proceedings)\cite{Coker2018}, we perform a survey to find optimal SPLC designs for a one dimensional representation of LUVOIR A. That is, the optimization is performed for an annular aperture with $R_\text{in}/R_\text{out}=0.1$, where $R_\text{in}$ and $R_\text{out}$ are the inner and outer radii, respectively. For each combination of focal plane mask and Lyot stop inner and outer radii, the optimization algorithm returns the apodizer that provides the highest throughput and a stellar PSF $<10^{-10}$ within the image of the focal plane mask opening in the final image plane over a spectral bandwidth of $\Delta\lambda/\lambda=0.1$. Using analytical models for the scientific yield of an exoplanet imaging mission from Stark et al. (2014)\cite{Stark2014}, we determine that the best SPLC in our survey (see Fig.~\ref{fig:dmsplcA}a-c) has focal plane mask inner and outer radii corresponding to 3.4~$\lambda/D_0$ and 26.4~$\lambda/D_0$, where $D_0=2R_\text{out}$. The Lyot stop has inner and outer radii of 0.24~$D_0$ and 0.78~$D_0$.


\subsubsection{DM-assisted apodization}

We insert the rotationally symmetric SPLC described above into the pupil of LUVOIR A such that the inscribed diameter $D_0=0.9D$, where $D$ is the full circumscribed diameter. The inner and outer radii of the focal plane mask correspond to 3.7~$\lambda/D$ and 29~$\lambda/D$. We optimized the DM shapes using FALCO to create the smallest possible stellar irradiance within the image of the focal plane mask opening computed at seven evenly spaced wavelengths over a spectral bandwidth of $\Delta\lambda/\lambda=0.1$. This was repeated for several strut widths ranging from 0.01\% to 1\% of $D$. 

\begin{figure}[p]
    \centering
    \includegraphics[width=\linewidth]{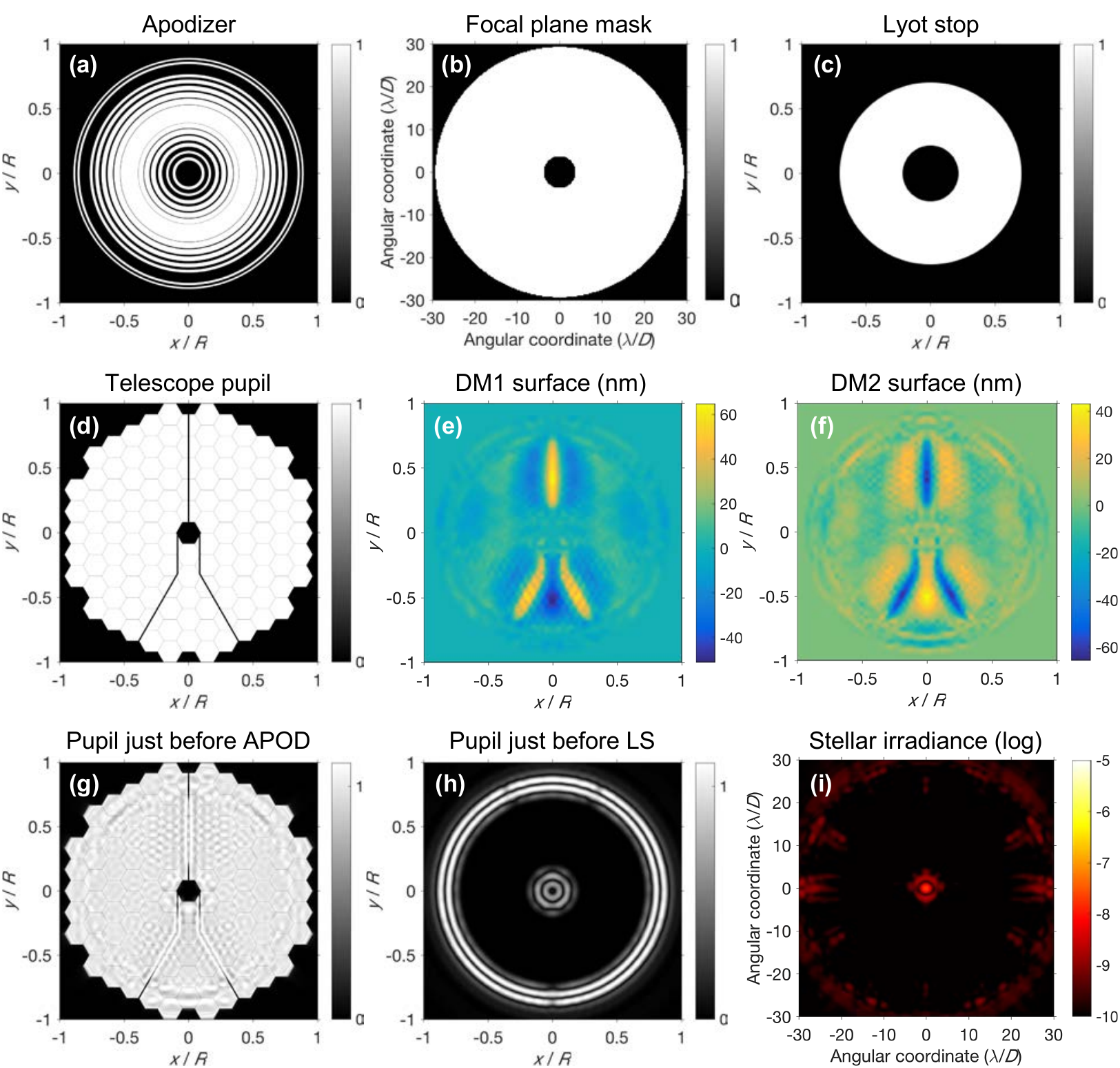}
    \caption{Shaped pupil Lyot coronagraph (SPLC) for LUVOIR A with DM-assisted apodization. (a)-(c) The 1-D radial solution with (a) a binary apodizer, (b) annular focal plane mask, and (c) annular Lyot stop. The struts and gaps between mirror segments in (d) the LUVOIR~A pupil are apodized by (e)-(f) the two DMs to create (g) the reshaped pupil. The binary apodizer removes the diffraction from the central obscuration such that (h) the stellar field is diffracted outside of the Lyot stop and doesn't propagate to (i) the final image plane. The strut width shown is 0.5\% of the pupil diameter.}
    \label{fig:dmsplcA}
\end{figure}

Figure~\ref{fig:dmsplcA} shows the resulting coronagraph design for a strut width of 0.5\% of $D$. The beam at the re-imaged telescope pupil upstream of the coronagraph masks (Fig.~\ref{fig:dmsplcA}d) is modified by the DMs (Fig.~\ref{fig:dmsplcA}e,f) to create the field amplitude shown in Fig.~\ref{fig:dmsplcA}g just prior to the apodizer (Fig.~\ref{fig:dmsplcA}a). After passing through the focal plane mask, the re-imaged pupil has most of the starlight (Fig.~\ref{fig:dmsplcA}h) diffracted outside of the Lyot stop (Fig.~\ref{fig:dmsplcA}c), where it is blocked before reaching the final image plane (Fig.~\ref{fig:dmsplcA}i).


\begin{figure}[p]
    \centering
    \includegraphics[width=\linewidth]{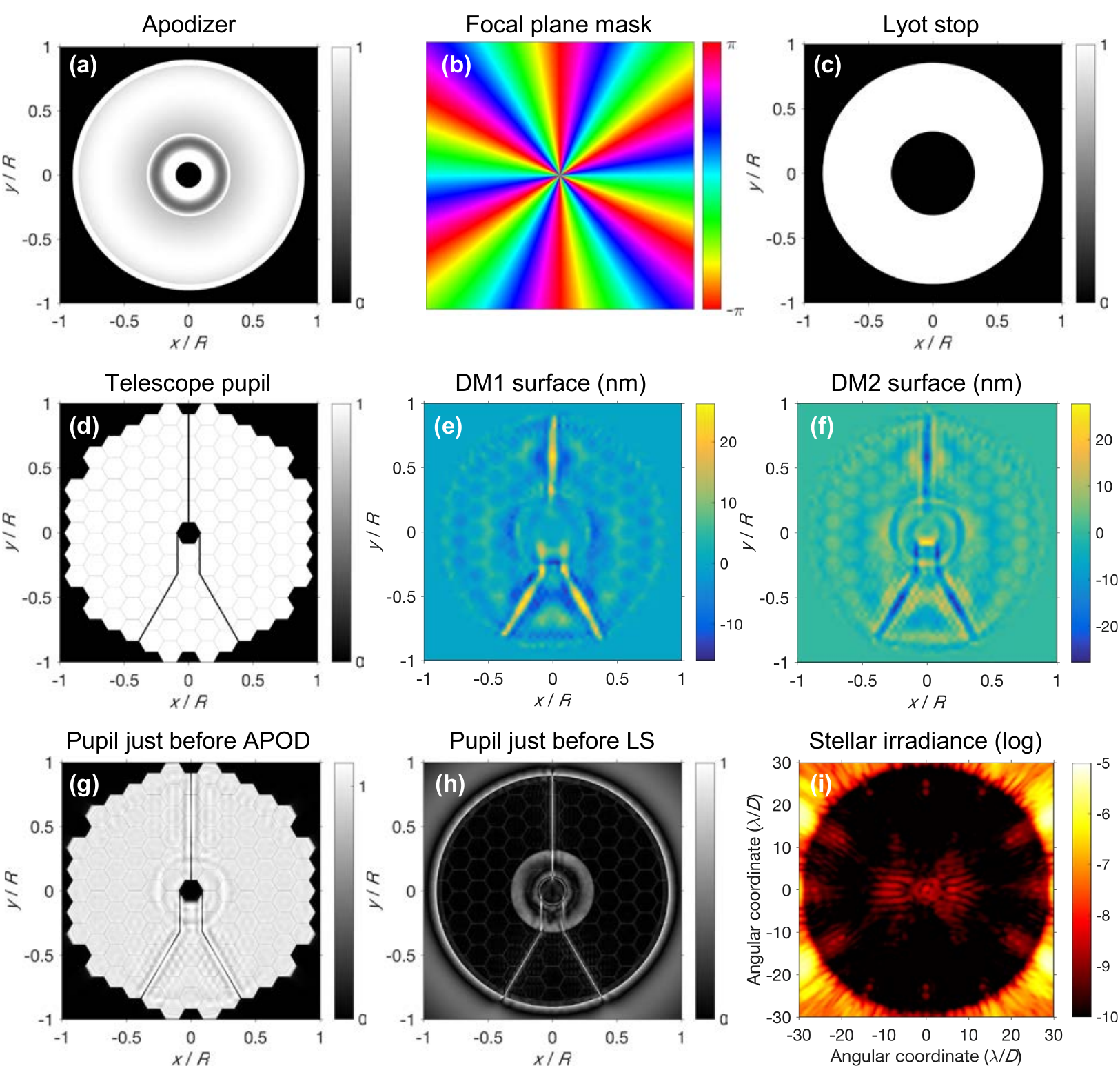}
    \caption{Vortex coronagraph for LUVOIR A with DM-assisted apodization. (a)-(c) The 1-D radial solution with (a) a grayscale apodizer, (b) a charge 6 vortex phase mask in the focal plane, and (c) annular Lyot stop. The struts and gaps between mirror segments in (d) the LUVOIR~A pupil are apodized by (e)-(f) the two DMs to create (g) the reshaped pupil. (h)~Some of the stellar field is diffracted outside of the Lyot stop. The starlight that propagates to (i) the final image plane is at high spatial frequency ($>$29 cycles per pupil diameter). The strut width shown is 0.5\% of the pupil diameter.}
    \label{fig:avcA}
\end{figure}

\subsection{Vortex coronagraph}

A vortex coronagraph\cite{Mawet2005,Foo2005,Ruane2018_JATIS} performs a similar optical function as the SPLC, but with very different coronagraph masks. Here, we use an apodized vortex coronagraph with a grayscale apodizer\cite{Mawet2013_ringapod,Fogarty2017,Ruane2016_SPIE,Ruane2017_SCDA,Zhang2018} that suppresses the diffraction from the central obscuration. The focal plane mask has complex transmittance given by $\exp(il\theta)$, where $l$ is an even integer known as the charge and $\theta$ is the azimuthal angle. We assume a charge six vortex phase mask ($l=6$) in the focal plane and an annular Lyot stop to take advantage of its innate robustness to low order aberrations\cite{Ruane2018_JATIS}. 

\subsubsection{1-D radial designs}

The modal symmetry of the vortex phase mask allows us to represent vortex coronagraphs with one dimensional functions. Ring apodizers\cite{Mawet2013_ringapod} and polynomial apodizers\cite{Fogarty2017} can be derived analytically. The apodizers may also be numerically optimized. As in the case of an SPLC, assuming a finite image plane mask yields a binary apodizer\cite{Carlotti2014}. Numerical solutions for grayscale apodizers are available for vortex masks with effectively infinite extent\cite{Ruane2016_SPIE,Ruane2017_SCDA,Zhang2018}.  

Again, we first optimize the coronagraph for the case of an annular aperture. The combination of the numerically optimized apodizer (Fig.~\ref{fig:avcA}a), charge six vortex phase mask in the focal plane (Fig.~\ref{fig:avcA}b), and annular Lyot stop (Fig.~\ref{fig:avcA}c) yields no diffraction up to a radius of 32~$\lambda/D$ over a theoretically infinite spectral bandwidth, where $D$ is the circumscribed diameter of the LUVOIR~A pupil. The inner and outer radii of the Lyot stop (0.32~$D$ and 0.86~$D$, respectively) maximize throughput.

\subsubsection{DM-assisted apodization}

As above, we start with the solution for the annular aperture and optimize the DM shapes to reduce diffraction from the struts and gaps between mirror segments in the LUVOIR~A pupil. We matched the outer radius of the image plane dark hole to that of the SPLC case (29~$\lambda/D$) and optimized for a spectral bandwidth of $\Delta\lambda/\lambda=0.1$ at seven discrete wavelengths. We also generated designs with $\Delta\lambda/\lambda=0.2$ at nine discrete wavelengths and a dark hole radius of 28~$\lambda/D$. Each were repeated for several strut widths ranging from 0.01\% to 1\% of $D$. 

Figure~\ref{fig:avcA} shows the apodized vortex coronagraph design for LUVOIR~A with a strut width of 0.5\% of $D$. The DMs (Fig.~\ref{fig:avcA}e,f) reshape the telescope pupil (Fig.~\ref{fig:avcA}d) to create the field amplitude shown in Fig.~\ref{fig:avcA}g just prior to the apodizer (Fig.~\ref{fig:avcA}a). After passing through the focal plane mask, some of the starlight (Fig.~\ref{fig:avcA}h) is diffracted outside of the Lyot stop (Fig.~\ref{fig:avcA}c), where it is blocked before reaching the final image plane (Fig.~\ref{fig:avcA}i). The signatures of the struts and gaps between mirror segmented are spatially high-pass filtered such that the starlight diffracts outside of 29~$\lambda/D$.

\begin{figure}[p]
    \centering
    \includegraphics[width=\linewidth]{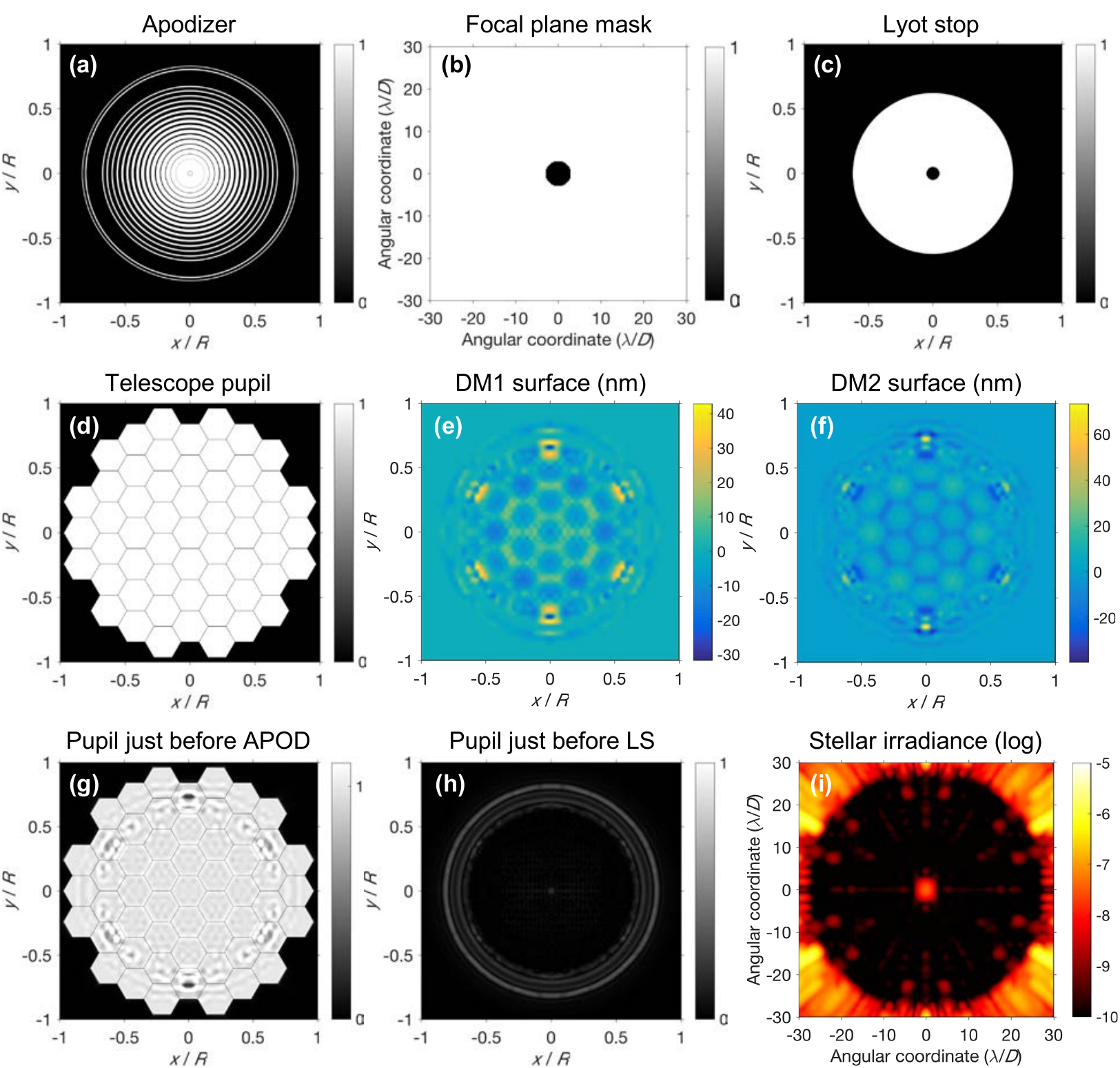}
    \caption{Same as Fig. \ref{fig:dmsplcA}, but for the LUVOIR~B pupil. The gap width is 0.1\% of the pupil diameter.}
    \label{fig:dmsplcB}
\end{figure}

\begin{figure}[p]
    \centering
    \includegraphics[width=\linewidth]{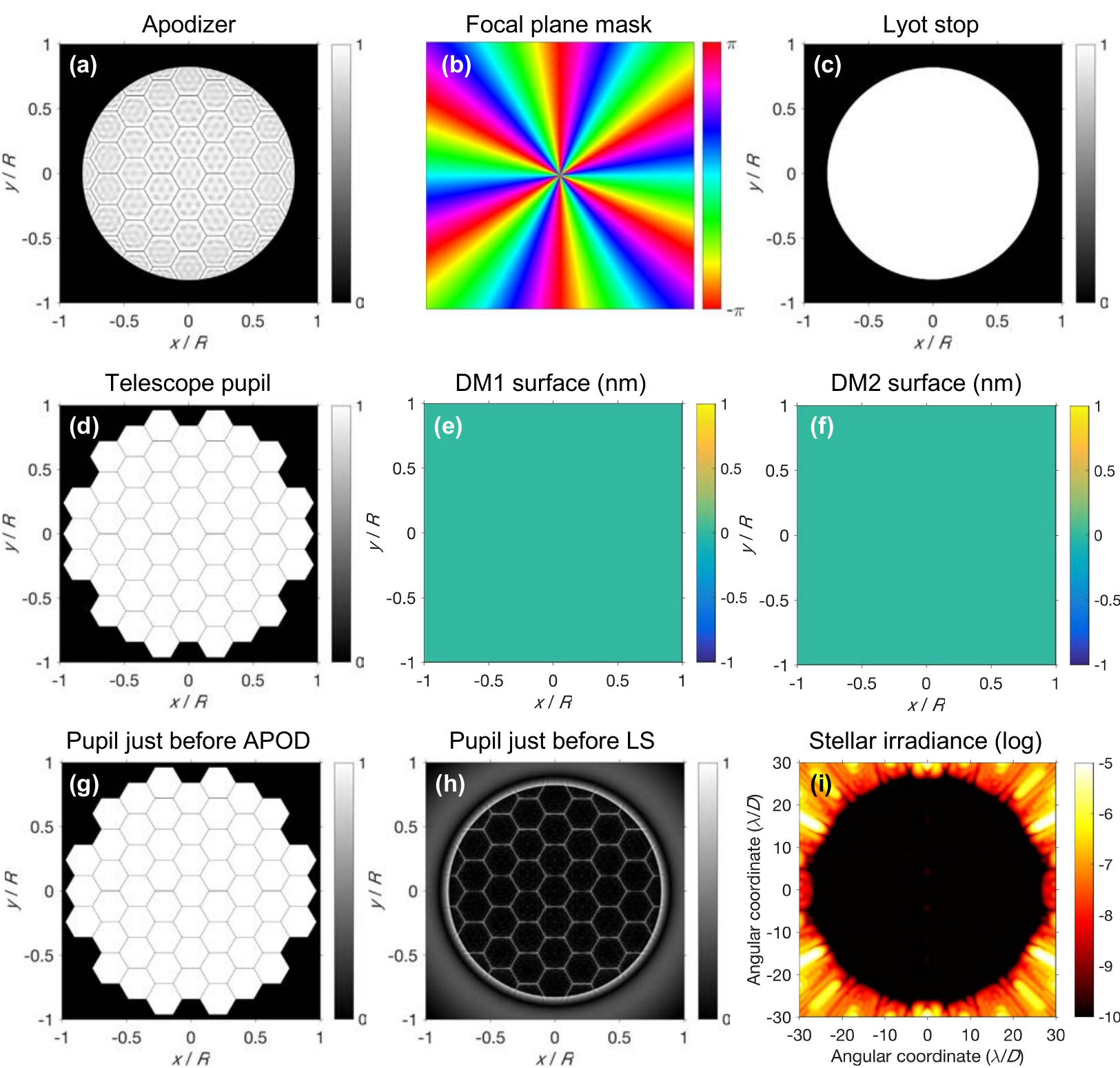}
    \caption{Vortex coronagraph for LUVOIR B with grayscale-only apodization. (a) The apodizer (diameter 0.84~$D$) is optimized to suppress diffraction from the gaps between mirror segments for (b) a charge 6 vortex phase mask in the intermediate focal plane and (c) a circular Lyot stop with diameter 0.82~$D$. (d) The telescope pupil is relayed through (e)-(f) flat DMs to create (g) an exact image of the pupil at the apodizer plane. The Lyot stop clips most of the (h) the starlight outside of the geometric pupil image and residual starlight diffracts outside of 28~$\lambda/D$. The gap width is 0.1\% of the pupil diameter.}
    \label{fig:gavcB}
\end{figure}

\begin{figure}[p]
    \centering
    \includegraphics[width=\linewidth]{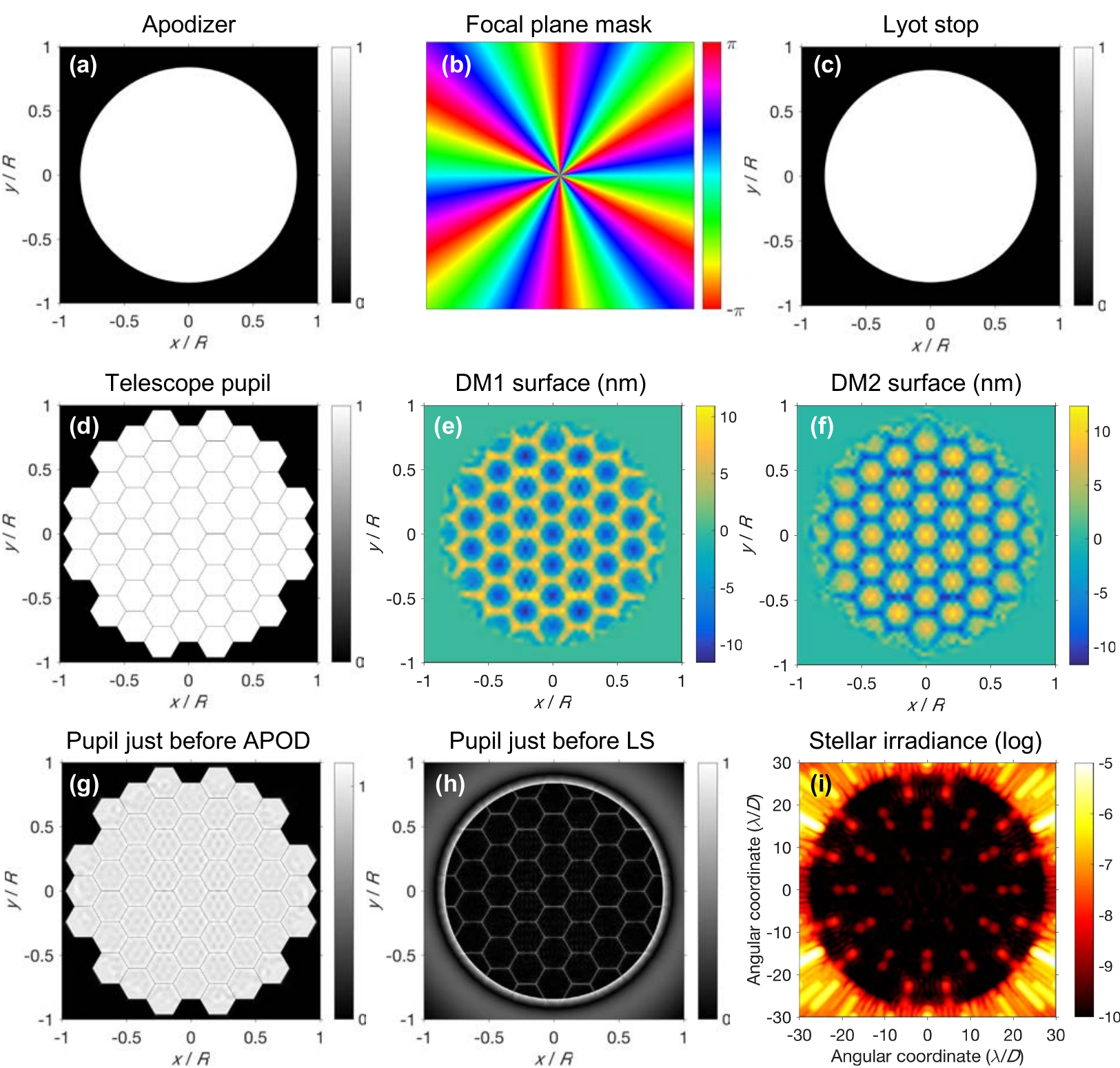}
    \caption{Vortex coronagraph for LUVOIR B with DM-only apodization. (a) A circular aperture (diameter 0.84~$D$) is inserted in the position of the apodizer to diffract most of the starlight with (b) a charge 6 vortex phase mask in the intermediate focal plane outside of (c) a circular Lyot stop with diameter 0.82~$D$. (d) The telescope pupil is reshaped by (e)-(f) the DMs to create (g) an apodized version of the pupil at the following pupil plane. (h) Starlight outside of the geometric pupil image is blocked by the Lyot stop and the residual leaked starlight diffracts outside of 28~$\lambda/D$. The gap width is 0.1\% of the pupil diameter.}
    \label{fig:avcB}
\end{figure}

\section{Designs for LUVOIR B}

In this section, we present shaped pupil Lyot and vortex coronagraph designs for the LUVOIR~B aperture (Fig.~\ref{fig:aperture}b). In each case, we compute the DM shapes needed to achieve a raw contrast of $<10^{-10}$ by suppressing the diffraction from the gaps between the mirror segments. 

\subsection{Shaped pupil Lyot coronagraph}

A shaped pupil Lyot coronagraph (SPLC) also needs a multi-ring binary apodizer to minimize diffraction from a circular aperture. Therefore, we first optimize the apodizer for a simple circular pupil, then compute the DM shapes afterward needed to create a dark hole with the segmented pupil. 

\subsubsection{1-D radial designs}

As in the case of LUVOIR A, we performed a design survey to find one dimensional SPLC designs by finding the apodizer that provides the highest throughput and a stellar PSF $<10^{-10}$ within the image of the focal plane mask opening in the final image plane over a spectral bandwidth of $\Delta\lambda/\lambda=0.1$ and $\Delta\lambda/\lambda=0.2$ as a function of focal plane mask and Lyot stop inner and outer radii. The best SPLC (see Fig. \ref{fig:dmsplcB}a-c) according to analytical yield models\cite{Stark2014} has a bandwidth of $\Delta\lambda/\lambda=0.2$, focal plane mask inner and outer radii corresponding to 2.5~$\lambda/D_0$ and 35~$\lambda/D_0$, where $D_0$ is the diameter of the circular aperture, and Lyot stop inner and outer radii of 0.06~$D_0$ and 0.74~$D_0$.

\subsubsection{DM-assisted apodization}

The diameter of the multi-ring binary apodizer is 84\% of the circumscribed diameter of the LUVOIR~B pupil. Therefore, the inner radius of the focal plane mask corresponds to 3~$\lambda/D$. We optimized the DM shapes to minimize the stellar irradiance with dark hole outer radius of 28~$\lambda/D$ at nine discrete wavelengths over a spectral bandwidth of $\Delta\lambda/\lambda=0.2$. This was repeated for several gap widths ranging from 0.01\% to 1\% of $D$. 

Figure~\ref{fig:dmsplcB} shows the resulting coronagraph design for a gap width of 0.1\% of $D$. The telescope pupil (Fig.~\ref{fig:dmsplcB}d) is modified by the DMs (Fig.~\ref{fig:dmsplcB}e,f) to create the apodized field amplitude shown in Fig.~\ref{fig:dmsplcB}g. The actuator strokes near the outer edge of the pupil are larger than usual. This feature will be investigated in future work. After the focal plane mask most of the starlight (Fig.~\ref{fig:dmsplcB}h) diffracted outside of the Lyot stop (Fig.~\ref{fig:dmsplcB}c), where it is blocked before reaching the final image plane (Fig.~\ref{fig:dmsplcB}i). Unlike the SPLC for LUVOIR~A, the focal plane mask and dark hole outer radius are not matched causing more leaked starlight outside of 28~$\lambda/D$.

\subsection{Vortex coronagraphs}

Vortex coronagraphs are well suited for off-axis telescopes as demonstrated in the context of the Habitable Exoplanet Observatory (HabEx) mission concept development\cite{Ruane2018_JATIS}. They differ from SPLC designs in that an apodizer is not required in the case of a circular, unobstructed pupil. However, defining a circular outer edge in the apodizer plane is beneficial. We present two varieties of vortex coronagraph designs that use grayscale-only and DM-only apodization. 

\subsubsection{Grayscale-only apodization}

We computed the grayscale apodization solution\cite{Ruane2017_SCDA} shown in Fig.~\ref{fig:gavcB}a using auxiliary field optimization (AFO)\cite{Jewell2017}. The outer radius of the apodizer (Fig.~\ref{fig:gavcB}a) and Lyot stop (Fig.~\ref{fig:gavcB}c) are 84\% and 82\% of the circumscribed diameter of the LUVOIR~B pupil (Fig.~\ref{fig:gavcB}d), respectively. Assuming the apodizer (Fig.~\ref{fig:gavcB}a) and vortex mask (Fig.~\ref{fig:gavcB}b) are perfectly achromatic, the DM surfaces (Fig.~\ref{fig:gavcB}e-f) remain flat, the pupil is unchanged before the apodizer (Fig.~\ref{fig:gavcB}d), and a very dark hole ($\ll10^{-10}$) at the final image plane (Fig.~\ref{fig:gavcB}i) is generated over an infinite spectral bandwidth out to a radius corresponding to 28~$\lambda/D$. The true limit of spectral bandwidth will be set by the wavefront sensing and correction of optical aberrations. 

\subsubsection{DM-only apodization}

We also present an alternate approach that uses the DMs to create the desired apodization pattern. In this case, we use a circular aperture at the apodizer position in Fig.~\ref{fig:layout} with diameter 84\% of the circumscribed diameter of the LUVOIR~B pupil (Fig.~\ref{fig:avcB}a). The focal plane mask (Fig.~\ref{fig:avcB}b) and Lyot stop (Fig.~\ref{fig:avcB}c) are the same as the previous case. The field at the re-imaged telescope pupil (Fig~\ref{fig:avcB}d) is modified by DMs (Fig~\ref{fig:avcB}e-f) to create an apodization pattern in the following pupil (Fig~\ref{fig:avcB}g). Visually the apodized pupil and the field at the Lyot stop (Fig~\ref{fig:avcB}h) appears similar to Fig.~\ref{fig:gavcB}a. However, since the phase applied by the DMs is fundamentally chromatic (i.e. scales as $1/\lambda$), starlight leaks inside the dark hole in the final focal plane (Fig~\ref{fig:avcB}i) at both ends of the spectral bandwidth ($\Delta\lambda/\lambda=0.2$) creating ``barbell" shaped speckles that generally increase in brightness toward the outer edge of the dark hole (28~$\lambda/D$). The centers of the barbells are located at the harmonic spatial frequencies associated with the segmentation pattern. 

\section{Performance comparisons}

In this section, we compare the performance of the five coronagraph designs presented above in terms of their raw contrast, throughput, and sensitivity to stellar size and low order aberrations. 

\subsection{EFC convergence}

\begin{figure}[t]
    \centering
    \includegraphics[width=\linewidth]{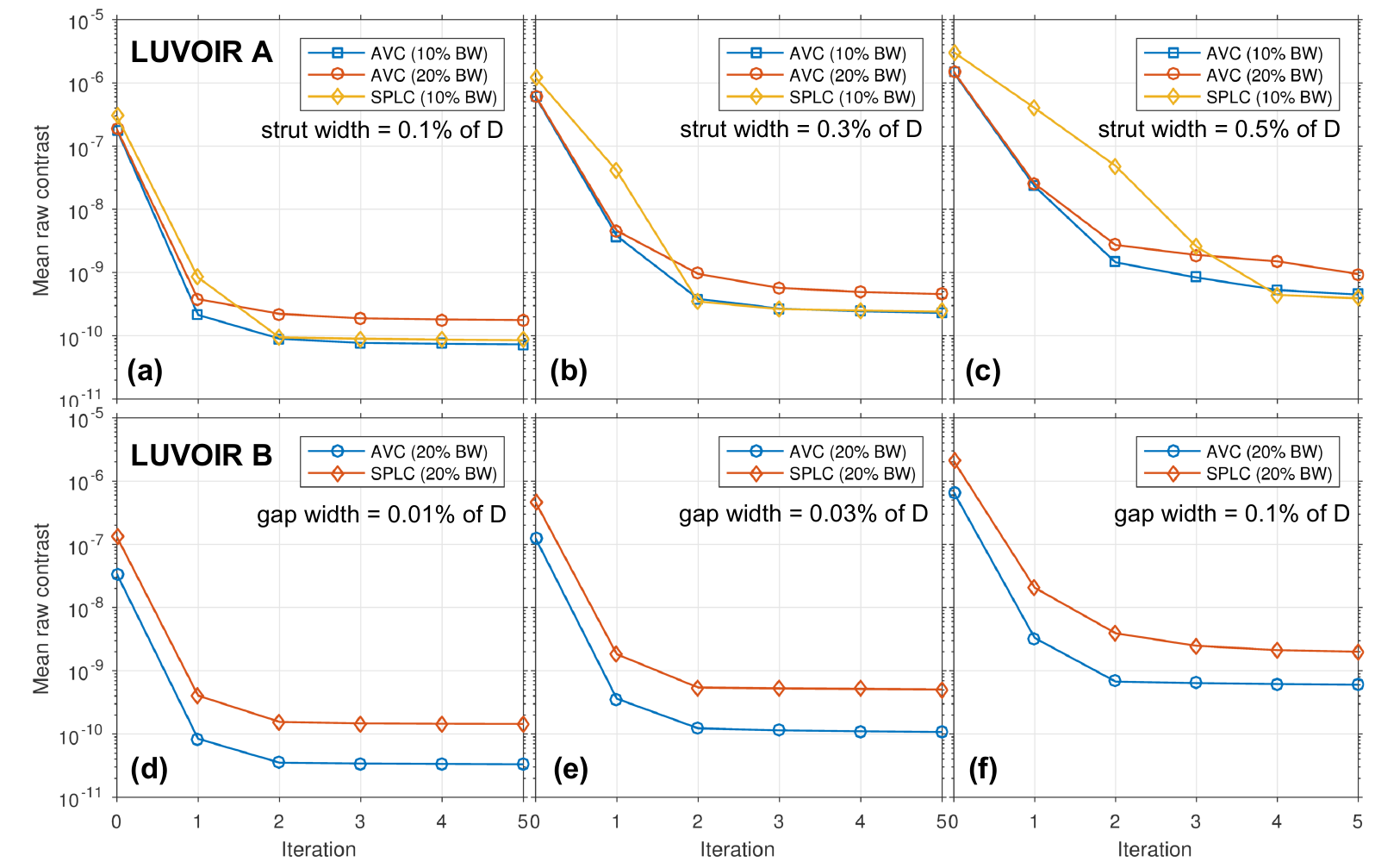}
    \caption{The mean raw contrast in the control region as a function of electric field conjugation (EFC) iteration and strut/gap size for (a)-(c) LUVOIR~A and (d)-(f) LUVOIR~B. We compare apodized vortex coronagraph (AVC) and shaped pupil Lyot coronagraph (SPLC) designs at various bandwidths (BW). }
    \label{fig:efc}
\end{figure}

The FALCO software package\cite{Riggs2018} implements electric field conjugation (EFC)\cite{Giveon2007} to compute the DM shapes that minimize the stellar irradiance in the dark hole using a linearized response matrix, also known as the control Jacobian, between the DMs and the electric field in the final focal plane. At each iteration, the DM is deformed by a linear step depending on the value of a regularization parameter. For this work, we re-compute the Jacobian and empirically determine the regularization parameter that yields the best raw contrast at each step. Alternate recipes may be easily explored with FALCO in future work. For most coronagraph designs, we reach the minimum raw contrast in the dark hole within five iterations. Figure~\ref{fig:efc} shows the mean raw contrast in the control region as a function of iteration and strut or gap size for the DM-apodized coronagraph designs presented above. In each case, the minimum raw contrast reached depends on the strut or gap size. 

For LUVOIR~A (Fig~\ref{fig:efc}a-c), a mean raw contrast of $10^{-10}$ is achieved for both the DM-apodized vortex coronagraphs (DM-AVC) and shaped pupil Lyot coronagraphs (DM-SPLC) with a strut size of 0.1\% of $D$. However, larger struts begin to limit the mean raw contrast to values greater than the typical goal of $10^{-10}$. The mean raw contrast of the AVC and SPLC are approximately the same for a spectral bandwidth of $\Delta\lambda/\lambda=0.1$ (10\%~BW), but the AVC raw contrast degrades by a factor of $\sim$2$\times$ when the spectral bandwidth increases to $\Delta\lambda/\lambda=0.2$ (20\%~BW). The same grayscale apodizer (Fig.~\ref{fig:avcA}) is used for both the 10\% and 20\% bandwidth cases because the apodizer optimization for vortex coronagraphs is inherently achromatic\cite{Ruane2016_SPIE}. 

For LUVOIR~B, on the other hand, the mean raw contrast (Fig~\ref{fig:efc}d-f) doesn't reach $10^{-10}$ if the gap size is greater than $\sim$0.01\% for an SPLC and $\sim$0.03\% for an AVC. However, unlike the case for the struts in the LUVOIR~A aperture, most of the diffraction from the inter-segment gaps appears near the outer edge of the control region ($>$15~$\lambda/D$ from the star) and therefore has a smaller effect on the scientific yield. Thus, DM-apodized coronagraphs can sufficiently suppress diffraction from gaps whose widths are on the order of $\sim$0.1\% of $D$.
Both the SPLC and AVC used in Fig~\ref{fig:efc}d-f are optimized for a spectral bandwidth of $\Delta\lambda/\lambda=0.2$. The grayscale-only vortex coronagraph (G-AVC) may be used instead of DM-AVCs for larger gap sizes where the EFC algorithm fails to converge to $10^{-10}$ over the desired region of the image plane. Future work will also investigate variations of the EFC control algorithm and regularization recipe to handle larger gap sizes. 

\subsection{Throughput}

\begin{figure}[t]
    \centering
    \includegraphics[width=0.85\linewidth]{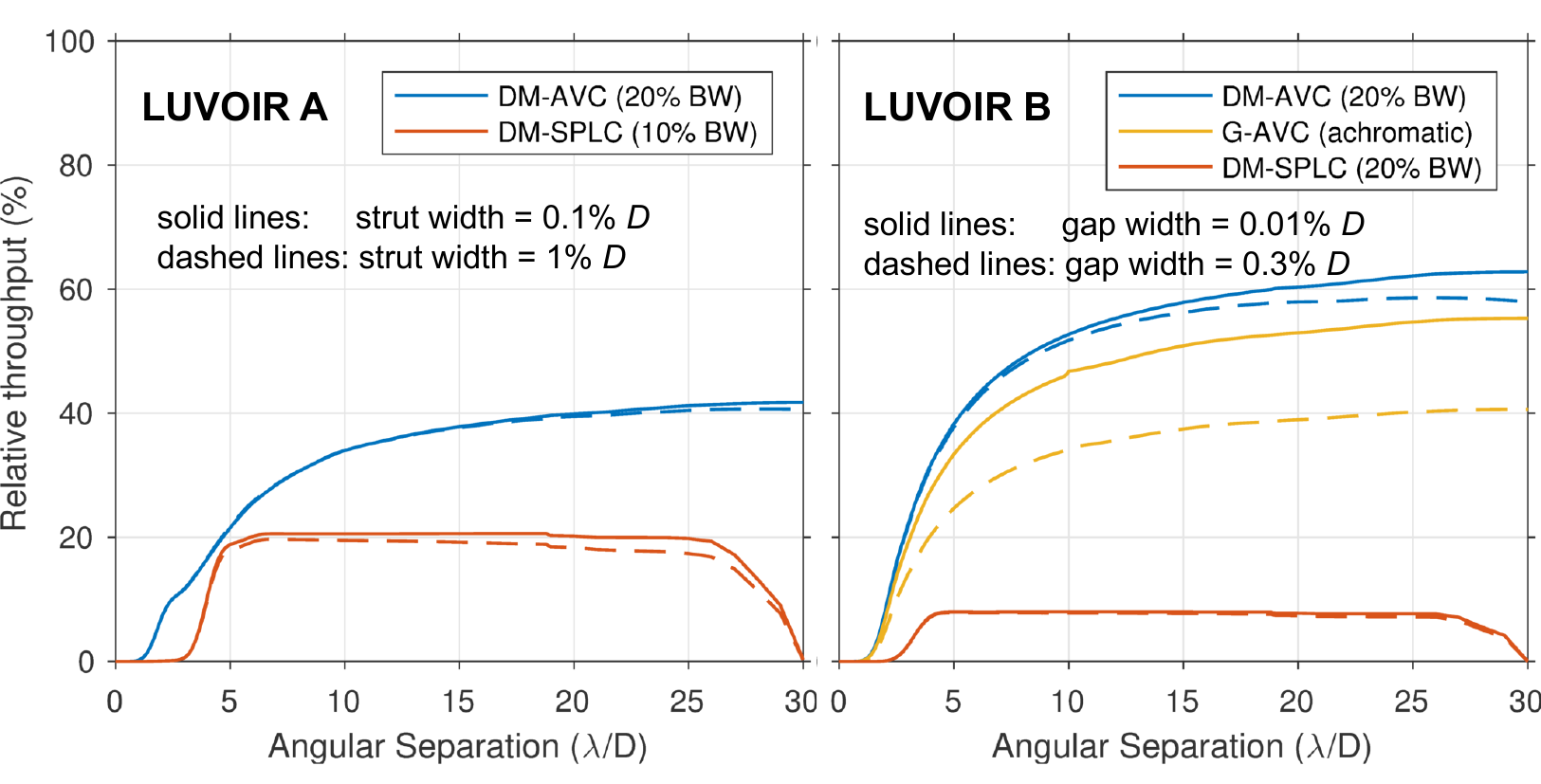}
    \caption{Relative throughput of a point source as a function of angular separation at the central wavelength for all five coronagraph designs. For LUVOIR~A (left), the solid and dashed lines correspond to strut widths of 0.1\% and 1\% of $D$, respectively. For LUVOIR~B (right), the solid and dashed lines correspond to gap widths of 0.01\% and 0.3\% of $D$, respectively. DM-apodized vortex coronagraphs (DM-AVCs) have the highest throughput in all cases. Grayscale apodized vortex coronagraphs (G-AVCs) may be used instead of DM-AVCs to improve raw contrast at the cost of throughput.}
    \label{fig:thpt}
\end{figure}

The relative throughput of a coronagraph compares the amount of light within the core of the off-axis PSF to the same quantity with the coronagraph masks removed and DMs flattened. Thus, this metric accounts for light that is blocked by apodizers, focal plane masks, Lyot stops as well as distortion to the off-axis PSF caused by those masks and the DMs. We integrate the signal in the PSF core within a radius of 0.7~$\lambda/D$. Figure~\ref{fig:thpt} shows the relative throughput of each coronagraph at the central wavelength. Overall, vortex coronagraphs have higher throughput and smaller inner working angles compared to SPLCs. The dashed lines show a minor loss in PSF quality due to an increase in DM stroke when the gaps become larger. The performance gained from using a vortex coronagraph is much more significant on the off-axis LUVOIR~B telescope. The DM-AVC yields the best throughput, but struggles to achieve a raw contrast better than $10^{-10}$ with gap sizes $>$0.1\% of $D$. For larger gaps sizes, it makes sense to switch to using a G-AVC to achieve better raw contrast. However, the throughput of a G-AVC is much more sensitive to the gap thickness, which also drives the telescope design to gaps on the order of $\sim$0.1\% of $D$ to take advantage of the high throughput of vortex coronagraphs. 

\subsection{Sensitivity to stellar size}

\begin{figure}[t]
    \centering
    \includegraphics[width=0.85\linewidth]{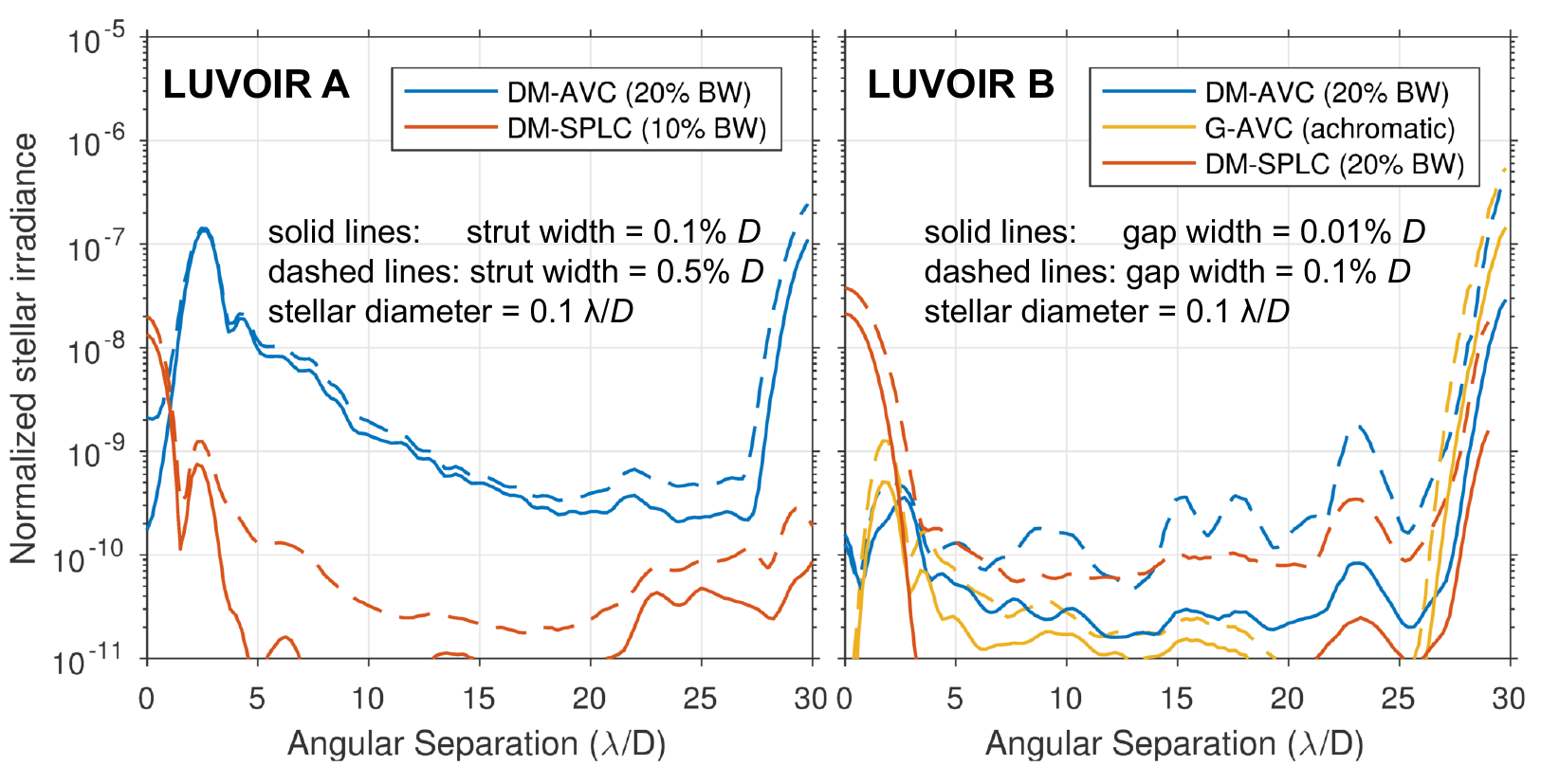}
    \caption{Azimuthal average of the stellar irradiance for a star with diameter of 0.1~$\lambda/D$ normalized to the peak of the PSF with the coronagraph masks removed and the DMs flat. The leaked starlight is slightly larger with thick struts or gaps (dashed lines) compared to thin ones (solid lines). The AVC is much more sensitivity to the stellar size on LUVOIR~A (i.e. a centrally obscured telescope) than LUVOIR~B (i.e. an off-axis telescope). The SPLC performs similarly in both cases. All of the coronagraph designs yield similar stellar irradiance levels on LUVOIR~B, except the performance degradation as a function of gap size is more apparent for the DM-apodized solutions. }
    \label{fig:stellarsize}
\end{figure}

The exquisite angular resolution of the LUVOIR telescope (potentially $\lesssim$10 milliarcseconds in the visible regime) make the finite size of stars a non-negligible source of noise. Figure~\ref{fig:stellarsize} shows the stellar irradiance for a star with diameter of 0.1~$\lambda/D$ ($\sim$1~milliarcsecond) normalized to the peak of the PSF with the coronagraph masks removed and the DMs flat. In the case of LUVOIR~A, the SPLC is much more robust to the size of the star than the AVC, which leaks $\sim$100$\times$ more starlight. However, in the case of LUVOIR~B, the AVC designs are nearly as robust as the SPLC design. This is a known effect due to diffraction from the central obscuration\cite{Ruane2017_SPIE}. The strut and gap sizes (solid vs. dashed lines) also influence the sensitivity to the size of the star, though not to the same extent as the central obscuration. Larger gaps sizes drive the DM stroke higher, which leads to higher sensitivity to the incoming angle of starlight with respect to the optical axis. The difference in phase shift applied by the DM on the beam originating from different points on the star increases with the stroke. This effect is also referred to as beam walk\cite{Noecker2005}. 

\subsection{Sensitivity to low order aberrations}

\begin{figure}[t]
    \centering
    \includegraphics[width=0.85\linewidth]{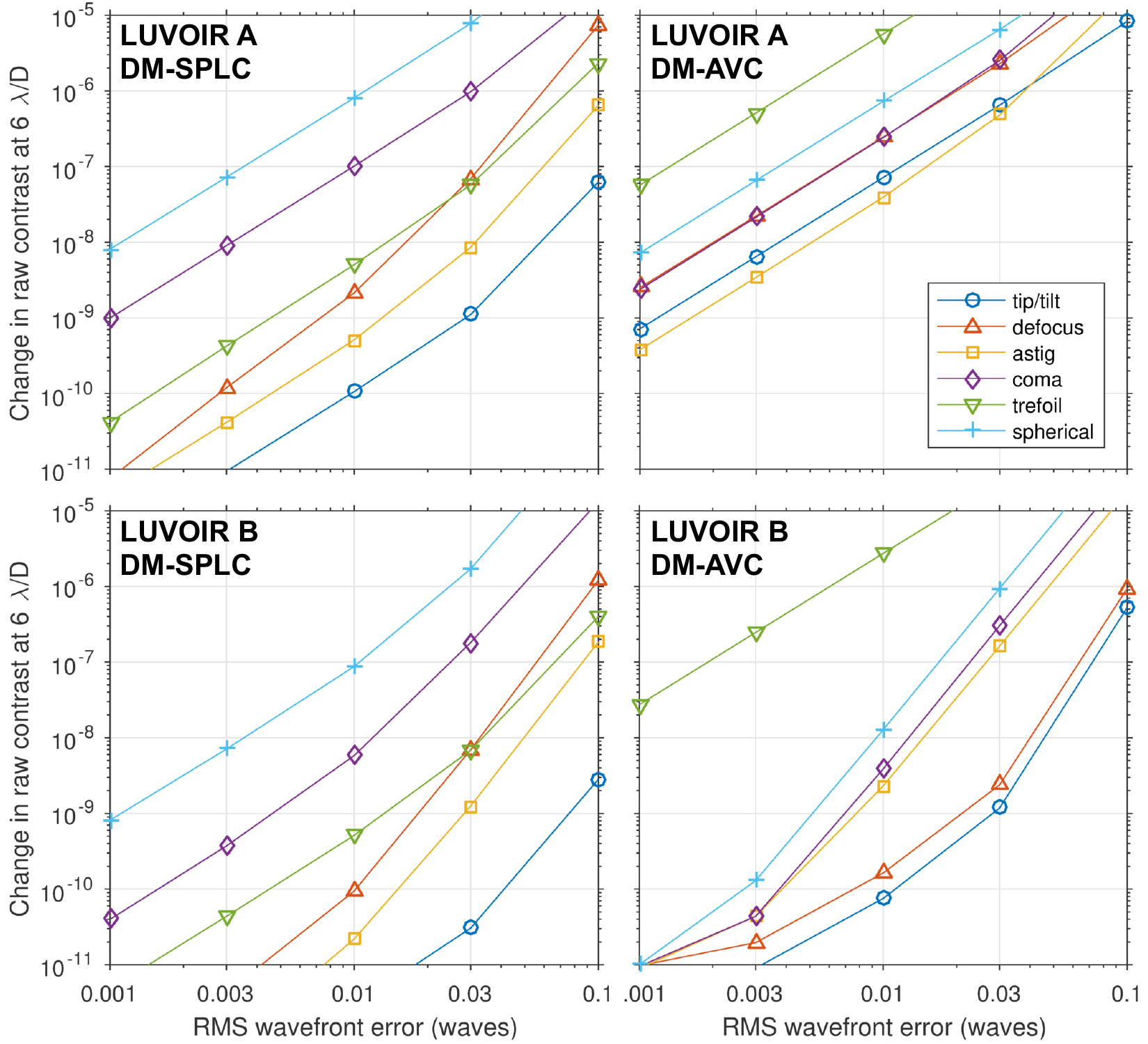}
    \caption{Sensitivity to low order aberrations. The change in raw contrast due to a wavefront error described by adding a Zernike polynomial to nominal wavefront at the entrance pupil of each coronagraph. The raw contrast is computed at the image plane coordinate corresponding to an angular separation at 6~$\lambda/D$ from the star. The LUVOIR~A pupil has a strut width of 0.1\% and the LUVOIR~B pupil has a gap size of 0.01\%. The strut and gaps sizes have a small influence on the sensitivity to low order aberrations.}
    \label{fig:Zsens}
\end{figure}

Each telescope aperture and coronagraph design combination has a unique response to low order aberrations. Figure~\ref{fig:Zsens} shows the change in raw contrast caused by introducing a wavefront error described by a Zernike polynomial at the plane of DM1. The raw contrast is computed at the image plane coordinate corresponding to an angular separation at 6~$\lambda/D$ from the star. Both coronagraph types are generally more robust to low order aberrations on LUVOIR~B than LUVOIR~A. The SPLC designs are more robust than the AVC on LUVOIR~A, but have similar sensitivity to coma and spherical aberrations. However, on LUVOIR~B, the AVC is much more robust to certain aberrations than the SPLC, such as coma and spherical, but is very sensitive to trefoil as expected for a charge 6 vortex coronagraph\cite{Ruane2018_JATIS}. Larger inner working angle SPLCs and higher charge AVCs (i.e. $l\ge8$) may be designed to be much more robust to low order aberrations than the examples shown here and thereby relax telescope and other instrumental stability requirements. 

\section{Conclusions}

We have made use of the new FALCO software package to design coronagraphs for LUVOIR~A and B and compare their performance. The DM-assisted apodization approach tends to yield higher throughput than conventional apodization with grayscale or binary amplitude masks, but struggles to suppress diffracted starlight when strut or gaps widths are too large (much greater than 0.1\%~of~$D$). Diffraction from thicker struts and/or gaps may be suppressed with conventional apodizers with degraded throughput (e.g. the G-AVC). AVCs have the advantage of providing higher throughput, larger bandwidth, and smaller effective inner working angles than SPLCs. However, SPLCs are generally more robust to stellar angular diameter and low order aberrations, especially on centrally-obscured telescopes. AVCs are very sensitive to stellar angular diameter with an centrally-obscured telescope (e.g. LUVOIR~A), which would severely limit their utility in that scenario, but are as robust as SPLCs on off-axis telescopes (e.g. LUVOIR~B). Overall, DM-AVCs on off-axis telescopes provide unrivaled all-around performance and the best scientific yield per unit telescope diameter and cost.


    

\acknowledgments  

G.~Ruane is supported by an NSF Astronomy and Astrophysics Postdoctoral Fellowship under award AST-1602444. This work was supported by the Exoplanet Exploration Program (ExEP), Jet Propulsion Laboratory, California Institute of Technology, under contract to NASA.



\bibliography{Library}   
\bibliographystyle{spiebib}   

\end{document}